\documentclass{webofc}\usepackage[varg]{txfonts} 
\begin{document}\title{Compact Exotic Tetraquark Mesons in
Large-{\boldmath$N_{\rm c}$} QCD}\author{Wolfgang Lucha\inst{1}
\fnsep\thanks{\email{Wolfgang.Lucha@oeaw.ac.at}}\and Dmitri
Melikhov\inst{1,2,3}\fnsep\thanks{\email{dmitri_melikhov@gmx.de}}
\and Hagop Sazdjian\inst{4}\fnsep
\thanks{\email{sazdjian@ipno.in2p3.fr}}}\institute{Institute for
High Energy Physics, Austrian Academy of Sciences,
Nikolsdorfergasse 18,\\A-1050 Vienna, Austria\and D.~V.~Skobeltsyn
Institute of Nuclear Physics, M.~V.~Lomonosov Moscow State
University,\\119991, Moscow, Russia\and Faculty of Physics,
University of Vienna, Boltzmanngasse 5, A-1090 Vienna, Austria
\and Institut de Physique Nucl\'eaire, CNRS-IN2P3, Universit\'e
Paris-Sud, Universit\'e Paris-Saclay,\\91406 Orsay Cedex, France}

\abstract{We embark on systematic explorations of the behaviour of
tetraquark mesons, \emph{i.e.}, colour-singlet bound states of two
quarks and two antiquarks,~in the (idealized) limit of a large
number of colour degrees of freedom, $N_{\rm c},$ of quantum
chromodynamics, QCD. Considering the scattering of two ordinary
mesons into two ordinary mesons, we start off with formulating a
set of selection criteria that should enable us to
\emph{unambiguously\/} single out precisely those contributions to
all encountered scattering amplitudes that potentially will
develop tetraquark poles. Assuming that tetraquark mesons do exist
and, if so, emerge in the contributions compatible with our
criteria at largest admissible order of $N_{\rm c},$ we deduce,
for the categories of tetraquarks that exhibit either four or only
two different open quark flavours, that the decay rates of these
tetraquark types are, at least, of order $1/N_{\rm c}^2$ and that
internal consistency requires all the members of the first
species~to exist pairwise, distinguishable by their favoured
two-ordinary-meson~decay channels.}\maketitle

\section{Stimulus: Decay Rates of Tetraquark States in the
{\boldmath$1/N_{\rm c}$} Expansion}The multiquark hadrons, such as
tetraquarks (mesonic bound states of two antiquarks and two
quarks) and pentaquarks (baryonic bound states of four quarks and
one antiquark), form, from both the theoretical and the
experimental points of view, one of the most challenging riddles
in hadron phenomenology. Starting, for obvious reasons, with the
mesonic and thus simpler case of tetraquarks, we approach this
issue by investigating the manifestation of tetraquark mesons as
intermediate states in scattering reactions of two ordinary mesons
into two ordinary mesons in form of contributions of
(narrow-width) poles to the associated scattering amplitudes
\cite{TQ1,TQ2,TQ3,TQ4,TQ5,TQ6}.

We tackle this task in two steps: First, we attempt to formulate
rigorous criteria that allow us to isolate the contributions of
Feynman diagrams capable of developing a pole interpretable as
being related to a compact (understood in contrast to a more
loosely bound molecular-type) tetraquark meson; we shall refer to
the members of this class of Feynman diagrams~potentially of
interest to us as the \emph{tetraquark-friendly\/}, or
\emph{tetraquark-phile\/}, ones. Then, we invoke QCD in some
well-defined disguise, to gain qualitative access to principal
features of selected variants of tetraquarks discriminated
basically by the flavour of their quark and antiquark
constituents.

\subsection{Identification of Tetraquark-phile Feynman Diagrams:
Set of Selection Criteria}\label{SC}In order to establish the
contribution of a tetraquark $T$ of mass $m_T$ (assumed not to
rise~with $N_{\rm c}$ in the limit $N_{\rm c}\to\infty$) viewed as
bound state of four (anti-) quarks
${{}^{\mbox{\tiny(}}_{}\hspace{-.17ex}\bar
q^{\mbox{\tiny)}}}\hspace{-.48ex}{}_{i}$ of masses $m_i$,
$i=1,\dots,4,$ to the $s$ channel of the scattering of two
appropriate ordinary mesons, of momenta $p$ and $q,$ we search for
a corresponding pole in four-point correlation functions of
quark-bilinear operators $j_{ij}\equiv\bar q_i\,q_j$ (dropping
parity and spin) interpolating the incoming and outgoing mesons
$M_{ij}$ (by having nonzero matrix elements between vacuum and
meson state), with decay constants~$f_{M_{ij}}$:\begin{equation}
\langle 0|j_{ij}|M_{ij}\rangle\equiv f_{M_{ij}}\ne0\ .\label{f}
\end{equation}We characterize a correlator contribution
potentially supporting a tetraquark pole (tagged~by a subscript T)
by the demand that the corresponding tetraquark-friendly Feynman
diagram~must\begin{itemize}\item involve the Mandelstam variable
$s\equiv(p+q)^2$ in a nontrivial (\emph{viz.}, non-polynomial)
way,~and\item admit an intermediate four-quark state, with a
branch cut starting at $s=(m_1+m_2+m_3+m_4)^2$.\end{itemize}The
existence of the latter threshold may be established by means of
the Landau equations~\cite{LDL}.

\subsection{Large-{\boldmath$N_{\rm c}$} QCD: Quantum
Chromodynamics' Limit {\boldmath$N_{\rm c}\to\infty$} \&
{\boldmath$1/N_{\rm c}$} Expansion}Large-$N_{\rm c}$ QCD
\cite{GH,EW} constitutes a rather extreme generalization of QCD
defined by letting the number $N_{\rm c}$ of the colour degrees of
freedom of QCD grow beyond bounds, \emph{i.e.}, by considering the
limit $N_{\rm c}\to\infty$ (and expansions thereabout), and
simultaneously demanding that the strong coupling $g_{\rm s}$
decreases for rising $N_{\rm c}$ such that the strong
fine-structure coupling $\alpha_{\rm s}$ behaves like$$\alpha_{\rm
s}\equiv\frac{g_{\rm s}^2}{4\pi}\propto\frac{1}{N_{\rm c}}\ .$$
Conventionally --- but not necessarily --- all quarks are required
to still transform according to the fundamental representation, of
dimension $N_{\rm c},$ of the then underlying gauge group ${\rm
SU}(N_{\rm c}).$ Large-$N_{\rm c}$ QCD provides an, at least,
qualitative understanding of crucial aspects of the hadron
spectrum within a, compared to QCD, simpler theoretical
environment and therefore has been applied \cite{SC,SW,KP,CL,MPR}
also to the issues of both existence and characterization of
tetraquark mesons. Among the first insights arising thereof is the
large-$N_{\rm c}$ behaviour of the decay constants $f_{M_{ij}}$~of
ordinary mesons, defined in Eq.~(\ref{f}), which prove to exhibit
a square-root increase with $N_{\rm c}$~\cite{EW}:
$$f_{M_{ij}}\propto\sqrt{N_{\rm c}}\ .$$

\subsection{Actual Target: Large-{\boldmath$N_{\rm c}$} Behaviour
of Total Decay Width of Tetraquark Mesons}The crucial feature of
any tetraquark meson is the $N_{\rm c}$ dependence of its total
decay width \cite{SW}. Experimental observability of a tetraquark
meson forbids its decay rate to grow with~rising $N_{\rm c},$ that
is to say, in the limit $N_{\rm c}\to\infty;$ consequently,
detectable tetraquarks must be narrow~states.

Since the two quarks and two antiquarks building up a given
tetraquark may be grouped to ordinary-meson states in two
different ways, in each case we analyze in parallel two scattering
channels discriminated according to whether the assignment of
quark flavours in the incoming and outgoing ordinary mesons is
identical (in that case dubbed as ``flavour-preserving'') or not
(in that case dubbed as ``flavour-rearranging,'' that is, has
undergone some flavour reshuffling).

Since, according to the above, we will have to deal with two
scattering channels, we better should be aware of the fact that it
might easily happen that the obtained consistency conditions
cannot be satisfied by a single tetraquark meson but inevitably
require the contributions of two tetraquarks (generally called
$T_A$ and $T_B$) with, of course, \emph{identical\/} quark-flavour
composition. We will take into account the possibility of
involuntary doubling of the tetraquark spectrum by suitable
formulation of the tetraquark-friendly four-point correlators from
the very beginning.

\section{Flavour-Exotic Tetraquark Mesons \boldmath{$\equiv$}
Four-Different-Flavour States}Genuinely flavour-exotic
tetraquarks, carrying not less than four different quark flavours,
may induce poles in two classes of four-point correlators related
to two-ordinary-meson scattering,\begin{subequations}\begin{align}
&\mbox{flavour preserving:}&&\hspace{-6ex}\langle
j^\dag_{12}\,j^\dag_{34}\,j_{12}\,j_{34}\rangle\ ,\qquad\langle
j^\dag_{14}\,j^\dag_{32}\,j_{14}\,j_{32}\rangle\ ;\qquad\label{ep}
\\&\mbox{flavour reshuffling:}&&\hspace{-6ex}\langle
j^\dag_{14}\,j^\dag_{32}\,j_{12}\,j_{34}\rangle\ .\label{er}
\end{align}\end{subequations}

\subsection{Flavour-Preserving Four-Current Correlation Functions:
\boldmath{$N_{\rm c}$}-Leading Diagrams}
\begin{figure}[h]
\centering\includegraphics[scale=.43668,clip]{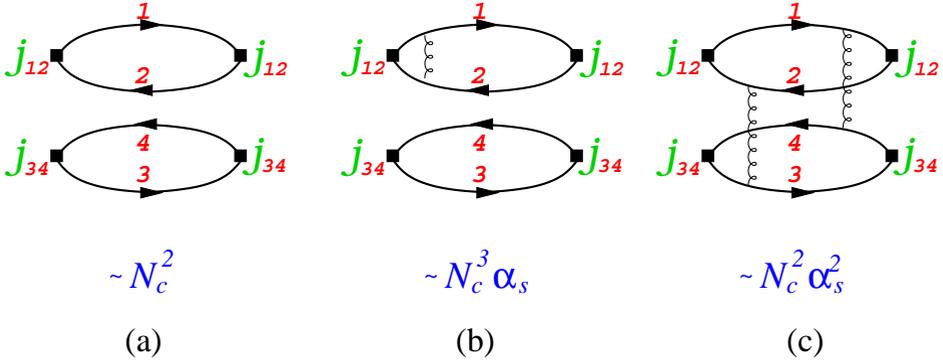}
\caption{Some of the contributions to the flavour-preserving
four-point Green functions (\ref{ep}) \cite[Fig.~1]{TQ1}.}
\label{Ep}\end{figure}

Figure \ref{Ep} depicts a
few typical examples of Feynman diagrams of high order in $N_{\rm
c}$ contributing to the flavour-preserving correlator $\langle
j^\dag_{12}\,j^\dag_{34}\,j_{12}\,j_{34}\rangle.$ Diagrams of the
type Fig.~\ref{Ep}(a) or \ref{Ep}(b) may be shown not to satisfy
the criteria of Subsect.~\ref{SC} and thus not to lead to
four-quark cuts~\cite{TQ1,TQ4}. In contrast to that, a Feynman
diagram of the type shown in Fig.~\ref{Ep}(c) might~develop
tetraquark poles. Accordingly, we find, as upper bounds to the
large-$N_{\rm c}$ behaviour of the tetraquark-phile contributions
to both flavour-preserving four-point correlation functions of
Eq.~(\ref{ep}) \cite{TQ1,TQ4},~that\begin{equation}\langle
j^\dag_{12}\,j^\dag_{34}\,j_{12}\,j_{34}\rangle_{\rm T}=O(N_{\rm
c}^0)\ ,\qquad\langle j^\dag_{14}\,j^\dag_{32}\,j_{14}\,j_{32}
\rangle_{\rm T}= O(N_{\rm c}^0)\ .\label{Np}\end{equation}The
contributions of the poles at $p^2=m^2_{T_{A,B}}$ to these
correlators involve ordinary-meson decay constants, $f_M,$ and
tetraquark--two-ordinary-meson transition amplitudes,
$A(M_{ij}\,M_{kl}\leftrightarrow T_{A,B})$:\begin{align*}\langle
j^\dag_{12}\,j^\dag_{34}\,j_{12}\,j_{34}\rangle_{\rm T}&=f_M^4
\left(\frac{|A(M_{12}\,M_{34}\leftrightarrow T_A)|^2}
{p^2-m^2_{T_A}}+\frac{|A(M_{12}\,M_{34}\leftrightarrow
T_B)|^2}{p^2-m^2_{T_B}}\right)+\cdots\ ,\\[1ex]\langle
j^\dag_{14}\,j^\dag_{32}\,j_{14}\,j_{32}\rangle_{\rm T}&=f_M^4
\left(\frac{|A(M_{14}\,M_{32}\leftrightarrow T_A)|^2}
{p^2-m^2_{T_A}}+\frac{|A(M_{14}\,M_{32}\leftrightarrow
T_B)|^2}{p^2-m^2_{T_B}}\right)+\cdots\ .\end{align*}

\subsection{Flavour-Rearranging Four-Current Correlation Functions:
\boldmath{$N_{\rm c}$}-Leading Graphs}
\begin{figure}[t]
\centering\includegraphics[scale=.4146,clip]{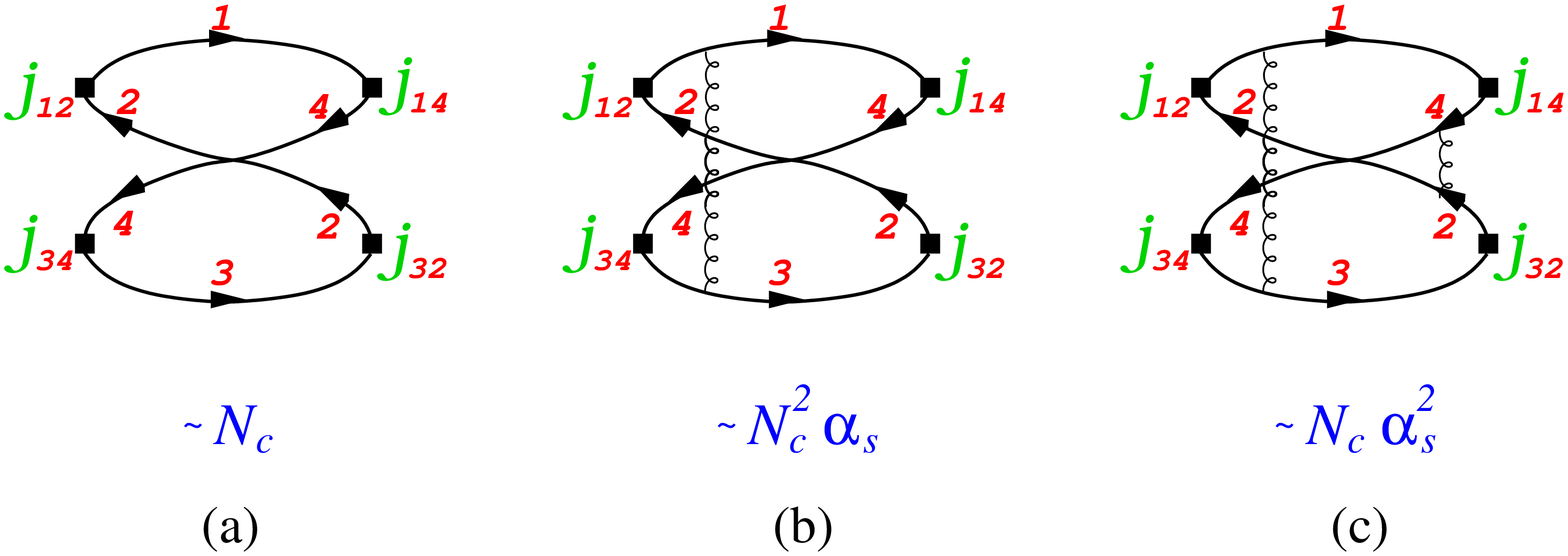}
\caption{Some of the contributions to the flavour-rearranging
four-point Green function (\ref{er}) \cite[Fig.~2]{TQ1}.}
\label{Er1}\end{figure}

Figure \ref{Er1} depicts
some representatives of Feynman diagrams of high order in $N_{\rm
c}$ that contribute to the flavour-rearranging correlator $\langle
j^\dag_{14}\,j^\dag_{32}\,j_{12}\,j_{34}\rangle.$ The diagrams of
the type Fig.~\ref{Er1}(a) or~\ref{Er1}(b) turn out to be not
compatible with the requirements proposed in Subsect.~\ref{SC}
\cite{TQ1,TQ4}. This result may be understood by unfolding the
quark lines to their respective box shape, as illustrated~by
Fig.~\ref{Er1} \cite{TQ3}. The decisive aspect in any such quest
for a tetraquark pole is the flow of colour~\cite{CL}.

\begin{figure}[b]
\centering\includegraphics[scale=.4146,clip]{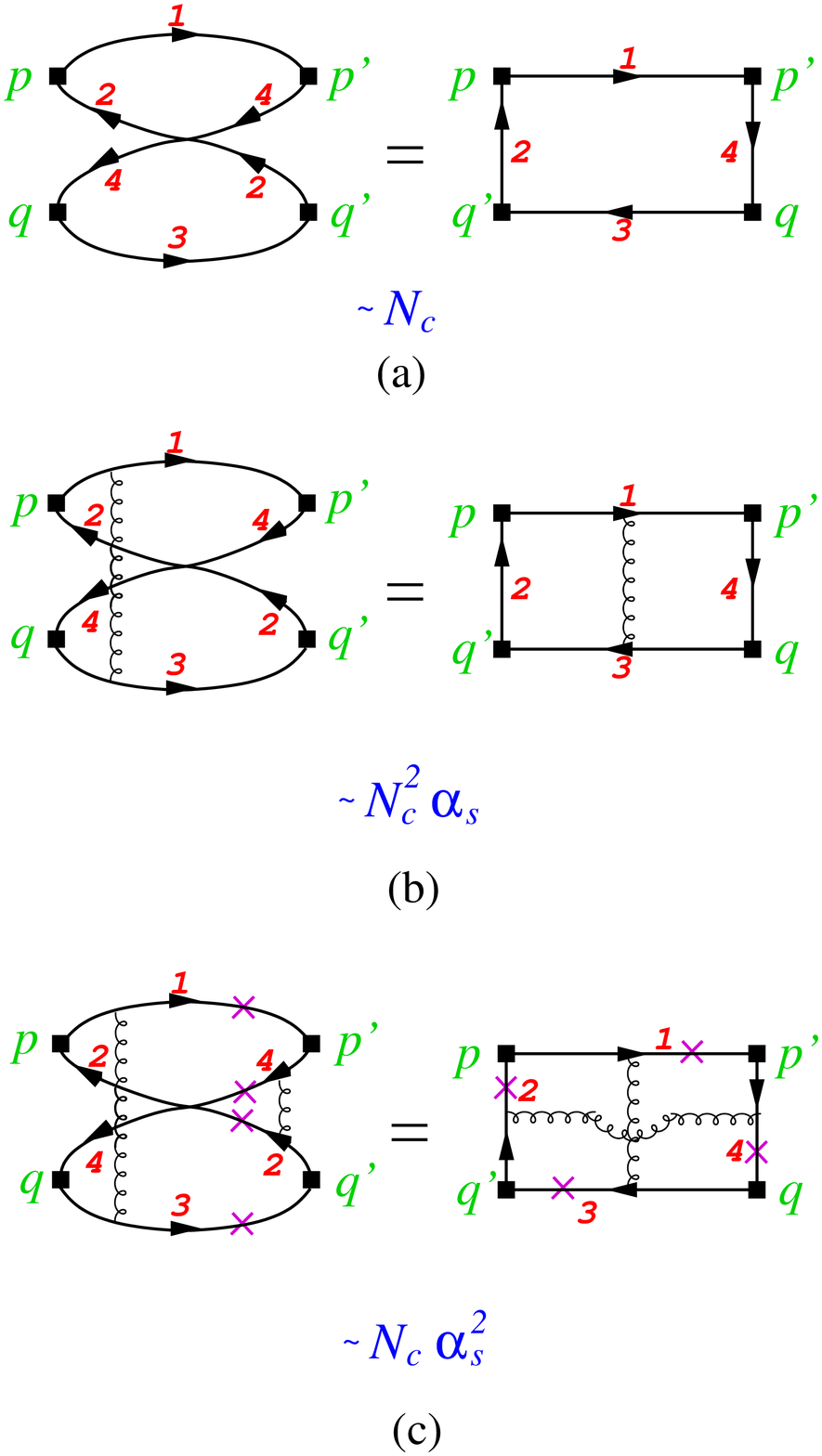}
\caption{Unfolding of quark lines to box shapes of all
illustrative Feynman diagrams depicted in Fig.~\ref{Er1}.}
\label{Er2}\end{figure}\clearpage

\noindent As a consequence, we infer, for the large-$N_{\rm c}$
behaviour of the tetraquark-phile contributions to the
flavour-rearranging four-point correlation function of type
Eq.~(\ref{er}), the upper bound~\cite{TQ1,TQ4}\begin{equation}
\langle j^\dag_{14}\,j^\dag_{32}\,j_{12}\,j_{34}\rangle_{\rm
T}=O(N_{\rm c}^{-1})\ .\label{Nr}\end{equation}Here, the required
tetraquark-induced redistribution of quark flavour at the poles at
$p^2=m^2_{T_{A,B}}$ combines different
tetraquark--two-ordinary-meson transition amplitudes
$A(M_{ij}\,M_{kl}\leftrightarrow T_{A,B})$:\begin{align*}\langle
j^\dag_{14}\,j^\dag_{32}\,j_{12}\,j_{34}\rangle_{\rm T}&=f_M^4
\left(\frac{A(M_{12}\,M_{34}\leftrightarrow T_A)\,
A(T_A\leftrightarrow M_{14}\,M_{32})}{p^2-m^2_{T_A}}\right.\\&
\hspace{3.75ex}+\left.\frac{A(M_{12}\,M_{34}\leftrightarrow T_B)\,
A(T_B\leftrightarrow M_{14}\,M_{32})}{p^2-m^2_{T_B}}\right)+\cdots
\ .\end{align*}

\subsection{Large-{\boldmath$N_{\rm c}$} Amplitudes of
Tetraquark--Two-Meson Transitions and Decay Widths}The \emph{upper
bounds\/} on the large-$N_{\rm c}$ dependence of the
tetraquark-friendly contributions to both categories of four-point
correlation function arising, on the one hand, from
flavour-preserving reactions, given by Eq.~(\ref{Np}), but, on the
other hand, from flavour-reshuffling reactions, given~by
Eq.~(\ref{Nr}), are certainly different. Let us, for definiteness,
assume that tetraquark poles contribute to such scattering
amplitude at the largest possible order of $N_{\rm c},$ being
tantamount to the lowest conceivable order of the perturbation
expansion of the scattering amplitude in powers of $1/N_{\rm c}.$

It is easy to convince oneself that in this case the arising
self-consistency conditions on the large-$N_{\rm c}$ dependence of
all the tetraquark--two-ordinary-meson transition amplitudes
entering in the involved pole terms cannot be satisfied by the
contribution of merely a single tetraquark. However, since the
$N_{\rm c}$ order of the flavour-rearranging scattering processes
is smaller than the $N_{\rm c}$ order of the flavour-preserving
scattering processes, the self-consistency conditions can be
satisfied by postulating the existence of two tetraquarks with,
clearly, identical flavour content but different couplings to that
two ordinary mesons with different assignment of quark~flavour:
\begin{subequations}\begin{align}A(T_A\leftrightarrow
M_{12}\,M_{34})&=O(N_{\rm c}^{-1})\ ,\qquad A(T_A\leftrightarrow
M_{14}\,M_{32})=O(N_{\rm c}^{-2})\ ,\label{A}\\
A(T_B\leftrightarrow M_{12}\,M_{34})&=O(N_{\rm c}^{-2})\ ,\qquad
A(T_B\leftrightarrow M_{14}\,M_{32})=O(N_{\rm c}^{-1})\ .\label{B}
\end{align}\end{subequations}Hence, the total decay rates of these
two tetraquarks $T_{A,B}$ exhibit the same large-$N_{\rm c}$
behaviour:$$\Gamma(T_A)=O(N_{\rm c}^{-2})\ ,\qquad\Gamma(T_B)=
O(N_{\rm c}^{-2})\ .$$

\subsection{Mixing of Flavour-Exotic Tetraquark States of Identical
Quark-Flavour Content}By construction, our two tetraquark states
$T_A$ and $T_B$ introduced above are bound states of~one and the
same set of (anti-) quarks and thus trivially exhibit the same
flavour quantum numbers. An upper bound to their mixing strength
$g_{AB}$ is provided by the flavour-rearranging correlator:
$$\langle j^\dag_{14}\,j^\dag_{32}\,j_{12}\,j_{34}\rangle_{\rm T}
=f_M^4\left(\frac{A(M_{12}\,M_{34}\leftrightarrow T_A)}
{p^2-m^2_{T_A}}\,g_{AB}\,\frac{A(T_B\leftrightarrow
M_{14}\,M_{32})} {p^2-m^2_{T_B}}\right)+\cdots\ .$$Our upper
bounds (\ref{A}) and (\ref{B}) to any \emph{$N_{\rm c}$-leading\/}
among the tetraquark--two-ordinary-meson transition amplitudes
$A(M_{ij}\,M_{kl}\leftrightarrow T_{A,B})$ translate into an upper
bound to the mixing strength:$$g_{AB}\le O(N_{\rm c}^{-1})\ .$$

\section{Flavour-Cryptoexotic Tetraquark Meson = Just Two Open
Flavours}For cryptoexotic tetraquarks $T=(\bar q_1\,q_2\,\bar
q_2\,q_3)$, with one quark--antiquark pair having the same
flavour, the following two classes of four-point Green functions
have to be taken into account:\pagebreak
\begin{subequations}\begin{align}&\mbox{flavour preserving:}&&
\hspace{-6ex}\langle
j^\dag_{12}\,j^\dag_{23}\,j_{12}\,j_{23}\rangle\ ,\qquad\langle
j^\dag_{13}\,j^\dag_{22}\,j_{13}\,j_{22}\rangle\ ;\qquad\label{cp}
\\&\mbox{flavour reshuffling:}&&\hspace{-6ex}\langle
j^\dag_{13}\,j^\dag_{22}\,j_{12}\,j_{23}\rangle\ .\label{cr}
\end{align}\end{subequations}

\subsection{Flavour-Preserving Four-Current Correlation Functions:
\boldmath{$N_{\rm c}$}-Leading Diagrams}
\begin{figure}[h]
\centering\includegraphics[scale=.43668,clip]{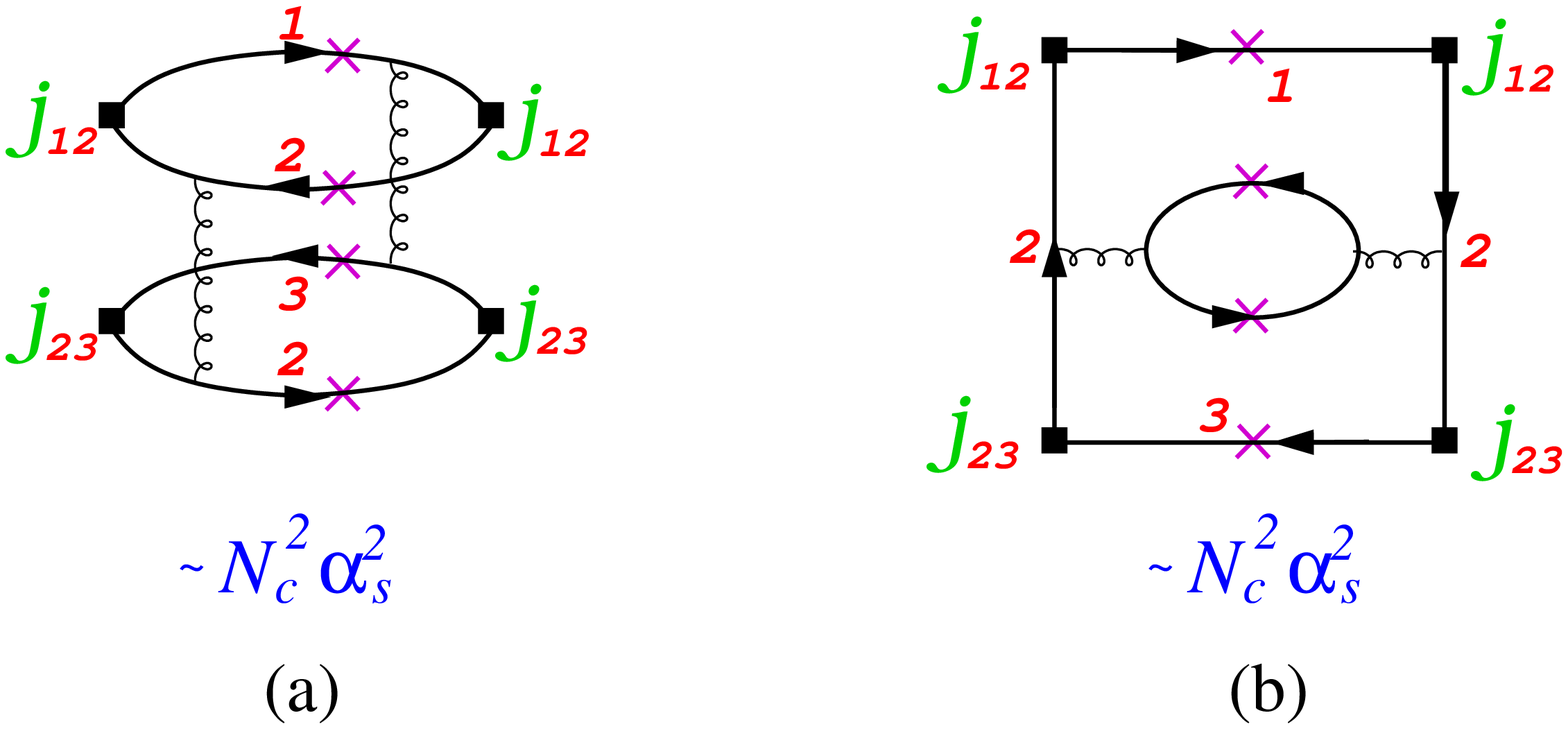}
\caption{Some of the contributions to the flavour-preserving
four-point Green functions (\ref{cp}) \cite[Fig.~3]{TQ1}.}
\label{Cp}\end{figure}

Figure \ref{Cp} depicts
two examples of $N_{\rm c}$-leading tetraquark-phile Feynman
diagrams contributing to the flavour-preserving correlator
$\langle j^\dag_{12}\,j^\dag_{23}\,j_{12}\,j_{23}\rangle,$ with
their tetraquark friendliness revealed by purple crosses. The
implied upper bounds to the large-$N_{\rm c}$ behaviour of these
correlators~are\begin{equation}\langle
j^\dag_{12}\,j^\dag_{23}\,j_{12}\,j_{23} \rangle_{\rm T}=O(N_{\rm
c}^0)\ ,\qquad\langle j^\dag_{13}\,
j^\dag_{22}\,j_{13}\,j_{22}\rangle_{\rm T}=O(N_{\rm c}^0)\
.\label{Npc}\end{equation}

\vspace{-3.579445ex}

\subsection{Flavour-Rearranging Four-Current Correlation Functions:
\boldmath{$N_{\rm c}$}-Leading Graphs}Figure \ref{Cp}(b)
demonstrates that the flavour-reshuffling Green function, $\langle
j^\dag_{13}\,j^\dag_{22}\,j_{12}\,j_{23}\rangle,$ receives
contributions of topologically new, and even $N_{\rm c}$-leading,
tetraquark-phile Feynman diagrams:\begin{equation}\langle
j^\dag_{13}\, j^\dag_{22}\,j_{12}\,j_{23}\rangle_{\rm T}=O(N_{\rm
c}^0)\ .\label{Nrc}\end{equation}

\vspace{-3.579445ex}\begin{figure}[h]
\centering\includegraphics[scale=.43668,clip]{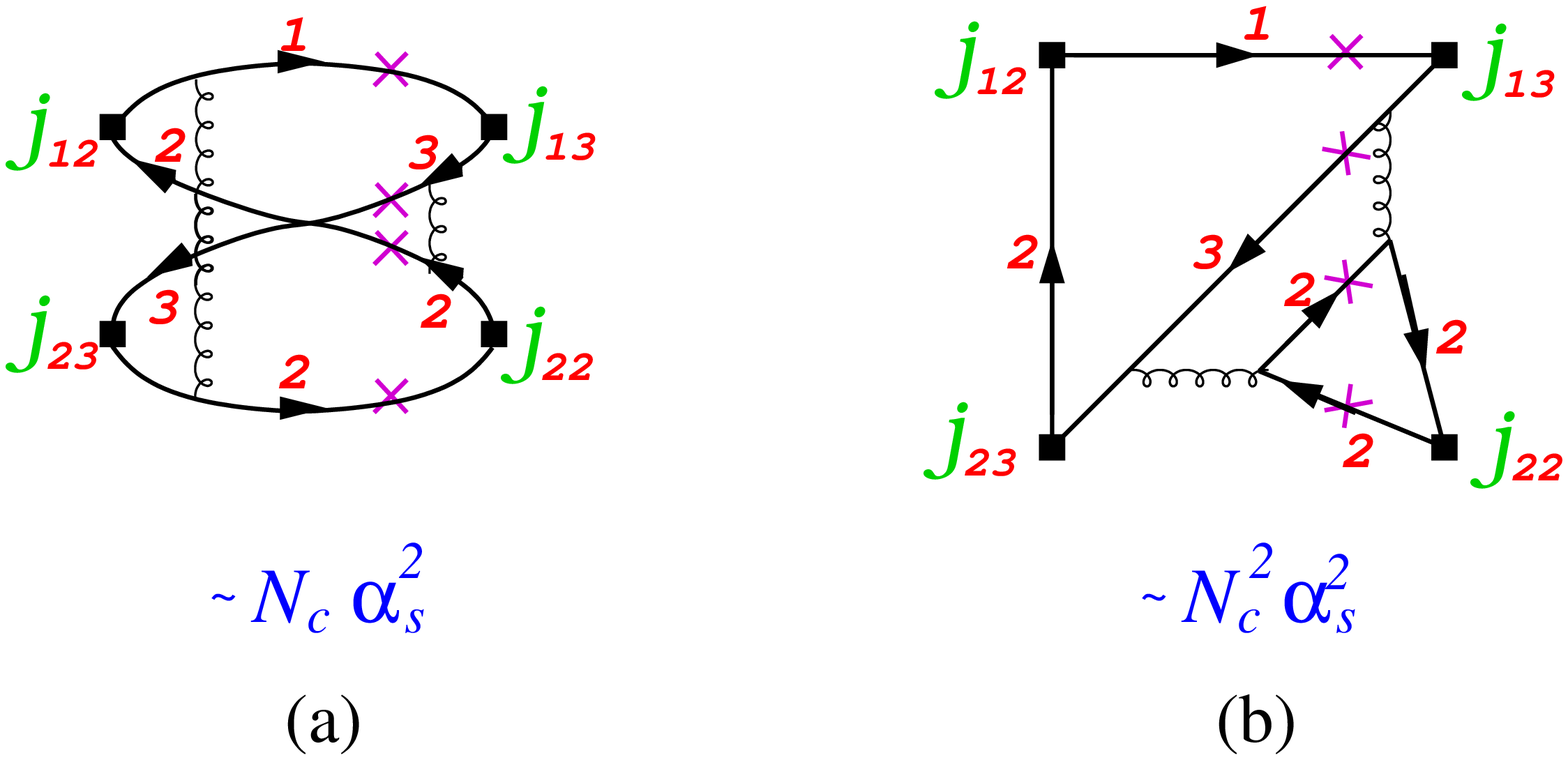}
\caption{Some of the contributions to the flavour-rearranging
four-point Green function (\ref{cr}) \cite[Fig.~4]{TQ1}.}
\label{Cr}\end{figure}

\subsection{Large-\boldmath{$N_{\rm c}$} Amplitudes of
Tetraquark--Two-Meson Transitions and Decay Widths}The \emph{upper
bounds\/} (\ref{Npc}) and (\ref{Nrc}) to the large-$N_{\rm c}$
dependence of the tetraquark-phile contributions to the
flavour-preserving and flavour-reshuffling four-point correlation
functions are identical. Thus, a single cryptoexotic tetraquark
$T$ can satisfy the emerging self-consistency conditions:
\begin{align*}A(T\leftrightarrow M_{12}\,M_{23})&=O(N_{\rm
c}^{-1})\ ,\qquad A(T\leftrightarrow M_{13}\,M_{22})=O(N_{\rm
c}^{-1})\\&\Longrightarrow\qquad\Gamma(T)=O(N_{\rm c}^{-2})\
.\end{align*}

\subsection{Mixing of Equal-Flavour Ordinary and
Flavour-Cryptoexotic Tetraquark Meson}Needless to say, each
cryptoexotic tetraquark $T=(\bar q_1\,q_2\,\bar q_2\,q_3)$ and
associated ordinary meson $M_{13}$ enjoy the same net
quark-flavour quantum numbers and can mix. An upper bound to their
mixing strength $g_{TM_{13}}$ can be estimated from, \emph{e.g.,}
the correlator $\langle j^\dag_{12}\,j^\dag_{23}\,j_{12}\,j_{23}
\rangle_{\rm T}$ according~to$$\langle j^\dag_{12}\,j^\dag_{23}\,
j_{12}\,j_{23}\rangle_{\rm T}=f_M^4\left(\frac{A(M_{12}\,M_{23}
\leftrightarrow T)}{p^2-m^2_T}\,g_{TM_{13}}\,\frac{A(M_{13}
\leftrightarrow M_{12}\,M_{23})}{p^2-m^2_{M_{13}}}\right)+\cdots\
.$$The three-ordinary-meson coupling, $A(M_{13}\leftrightarrow
M_{12}\,M_{23}),$ has been found \cite{GH,EW} to behave~like
$$A(M_{13}\leftrightarrow M_{12}\,M_{23})\propto\frac{1}
{\sqrt{N_{\rm c}}}\ .$$Thus, the large-$N_{\rm c}$ behaviour of
any cryptoexotic-tetraquark--ordinary-meson mixing satisfies
$$g_{TM_{13}}\le O(N_{\rm c}^{-1/2})\ ,$$if, for definiteness, we
choose to adhere to the upper bound $A(T\leftrightarrow
M_{12}\,M_{23})=O(N_{\rm c}^{-1})$ to the large-$N_{\rm c}$
behaviour of the cryptoexotic-tetraquark--two-ordinary-meson
transition amplitudes.

\section*{Acknowledgements}D.~M.\ is grateful for support by the
Austrian Science Fund (FWF) under project P29028-N27.

\end{document}